\documentclass[twocolumn,prl,10pt,amssymb,amsmath,amsfonts]{revtex4}
\usepackage{graphicx}
\usepackage{mathrsfs}
\usepackage{bm}
\usepackage{wasysym}    
\usepackage{color}
\def\be{\begin{equation}}
\def\ee{\end{equation}}
\def\bc{\begin{center}}
\def\ec{\end{center}}
\newcommand{\dagga}{{\phantom{\dagger}}}
\newcommand{\bea}{\begin{eqnarray}}
\newcommand{\eea}{\end{eqnarray}}
\usepackage[colorlinks,bookmarks=false,citecolor=red,linkcolor=red,urlcolor=blue]{hyperref}
\begin{document}
\title{Sliding phase in randomly stacked 2D superfluids/superconductors}
\author{\href{http://www.lpt.ups-tlse.fr/spip.php?article53&lang=en}{Nicolas Laflorencie}}
\email{laflo@irsamc.ups-tlse.fr}
\address{Laboratoire de Physique Th\'eorique, Universit\'e de Toulouse, UPS, (IRSAMC), F-31062 Toulouse, France}
\date{~}
\begin{abstract}
{Using large scale quantum Monte Carlo simulations of lattice bosonic models, we precisely investigate the effect of weak Josephson tunneling between 2D superfluid or superconducting layers. In the clean case, the Kosterlitz-Thouless transition immediately turns into 3DXY, with phase coherence and superflow in all spatial directions, and a strong enhancement of the critical temperature. However, when disorder is present, rare regions fluctuations can lead to an intermediate finite temperature phase --- the so called sliding regime --- where only 2D superflow occurs within the layers without any transverse superfluid coherence, while a true 3D Bose-Einstein condensate exists.
Critical properties of such an unconventional regime are carefully investigated.\\}
\end{abstract}
\maketitle
{\it{Introduction---}}
In condensed matter quantum systems, disorder is known to lead to a large variety of fascinating  phenomena. Relevant in strictly one-dimensional (1D) systems~\cite{Giamarchi87-Fisher95}, the situation is usually less universal in higher dimension~\cite{Miranda05-Mirlin08}. In that respect, the case of disordered superfluids or superconductors has been intensively studied during the past decades~\cite{Fisher89-Wallin94-Imada98}. More recently, ultracold atom experiments have triggered increasing interest for disordered induced localized phases of interacting bosonic systems~\cite{Fallani07-Lugan07-Deissler10}. 
In such a context, the question of the competition between superfluidity or superconductivity and disorder has led to a large amount of fascinating works focusing on bosonic quantum glasses. In particular, Bose glass physics has been addressed in various condensed matter systems: superconductors, cold atoms, polaritons, quantum antiferromagnets~\cite{Beek95-Nohadani05-Malpuech07-Muller09-Delande09-Gurarie09-Hong10-Lin11}. 
Among all these recent developments, it appears that the subtle interplay between disorder and dimensional crossover has been overlooked, despite the strong experimental interest in layered systems such as superconducting cuprates~\cite{Lee06-Benfatto07}, thin superfluid films~\cite{Minnhagen87}, or quasi-2D bosonic gases~\cite{Hadzibabic06}. Only recently, two simultaneous analytical works~\cite{Mohan10,Pekker10} have predicted for weakly disordered coupled superfluid layers an intermediate sliding regime (first evoked in the context of DNA complexes~\cite{Ohern99}), where there is no phase locking between 2D superfluids. Such exotic sliding superfluid state had also been pursued in the context of the frustrated antiferromagnetic layered system BaCuSi$_2$O$_6$~\cite{Sebastian06-Batista07}. However, frustration-induced 2D classical decoupling does not survive quantum fluctuations which restore 3D superfluidity~\cite{Yildirim96-Maltseva05,Laflorencie09-11}. Dynamical decoupling of superconducting layers induced by a stripe order has also been  discussed for La$_{2-x}$Ba$_x$CuO$_4$~\cite{Berg07}.\\

In this work we focus on weakly disordered superfluid or superconducting layers. To this aim, we use large scale quantum Monte Carlo (QMC) simulations of lattice bosonic models and test predictions of Refs.~\cite{Mohan10,Pekker10} regarding the occurrence of a stable sliding regime for disordered coupled superfluid layers. We first present evidences that when isolated layers are coupled by a Josephson tunneling $t_\perp$, a finite transverse stiffness 
immediately develops for any $t_\perp\neq 0$. On the other hand, when layers of  two types, having different individual Kosterlitz-Thouless temperatures, are randomly stacked and 3D coupled, QMC simulations 
show evidences for a wide temperature window where a sliding phase is achieved. We then present a detailed study of the critical properties of such a state, and discuss momentum space properties.\\

{\it{Coupled clean superfluid layers---}}
\begin{figure*}[!t]
\centering
\includegraphics[clip,width=2\columnwidth]{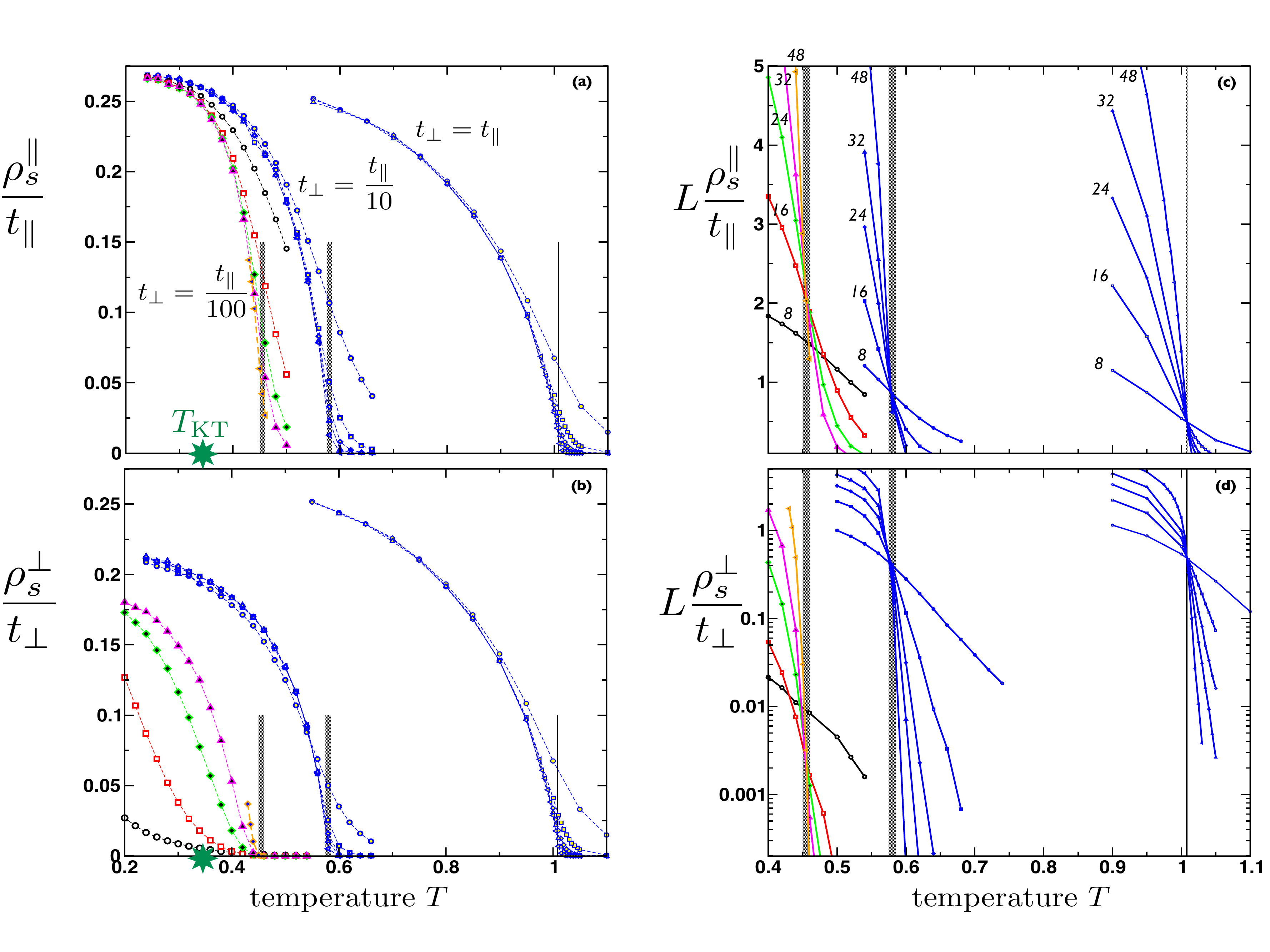}
\caption{(color online) QMC results for the 3DXY transition of clean coupled superfluid or superconducting layers of hard-core bosons [Eqs.~\ref{eq:Hhcb1} and \ref{eq:Hhcb2}].
In-plane and out-of-plane superfluid stiffnesses are shown in panels (a-b) for three representative values of $t_\perp/t_\parallel$, and different system sizes: $L=8~(circle),~ L=16~(square),~L=24~(diamond),~L=32~(up~triangle),~L=48~left~triangle)$. In panels (c-d), the same QMC data multiplied by $L$ display a crossing at the critical temperature $T_c$.}
\label{fig:clean}
\end{figure*}
One of the simplest lattice model for studying 2D superfluidity is the Bose-Hubbard Hamiltonian on a $L \times L$ square lattice 
\bea
{\cal H}&=&-\frac{t_\parallel}{2}\sum_{\langle i j\rangle}\left(b^{\dagger}_{i}b_{j}^{\dagga}+b^{\dagger}_{j}b_{i}^{\dagga}\right)\nonumber\\
&-&\mu\sum_i n_i+U\sum_in_i(n_i-1).
\label{eq:Hhcb1}
\eea
Particle hopping is restricted to pairs $\langle ij\rangle$ of neighboring sites, and the density of bosons $\langle n\rangle =\sum_{i}\langle b_i^{\dagger}b_i^\dagga\rangle/L^2$ is controlled by the chemical potential $\mu$. In the following, we will work in the limit of infinite on-site repulsion $U\to\infty$: the hard-core bosons limit, know as the Tonk-Girardeau regime, as achieved in cold atom experiments~\cite{Paredes04}. Ground-state and finite temperature properties of 2D hard-core bosons on a square lattice are pretty well-known~\cite{Hebert02-Bernardet02-Coletta12}. For $|\mu|<2t_\parallel$, the system is superfluid {\it{and}} Bose-condensed at $T=0$ whereas at finite temperature, 
only superfluidity survives~\cite{Mermin66-Hohenberg67} below a finite Kosterlitz-Thouless (KT) temperature $T_{\rm KT}$ where the superfluid density $\rho_{\rm sf}$ displays a universal jump~\cite{Harada97}. \\

\begin{figure}[!hb]
\centering
\includegraphics[clip,width=\columnwidth]{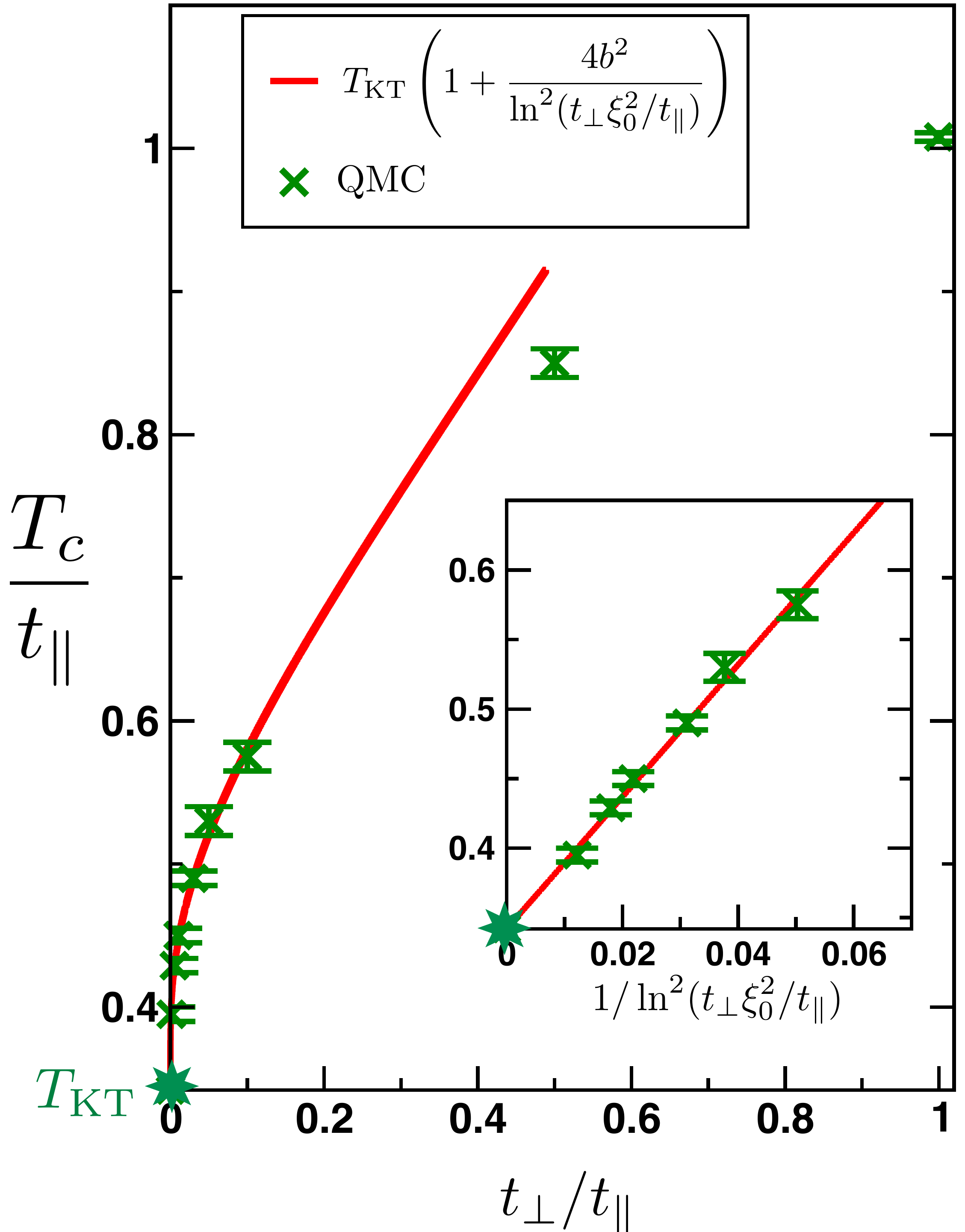}
\caption{{{(color online) Behavior of the 3DXY transition temperature $T_c/t_{\parallel}$ of clean superluid layers, plotted against interlayer hopping $t_\perp/t_\parallel$. Green crosses are the QMC estimates obtained from crossing analysis displayed in Fig.~\ref{fig:clean} and the red line is a fit to the form Eq.~\ref{eq:Tc} using $T_{\rm KT}=0.342 t_\parallel$~\cite{Harada97} (green star), $b=1.86$, and $\xi_0=0.34$.}}}
\label{fig:PHDGCLEAN}
\end{figure}

When superfluid layers are Josephson coupled by
\be
{\cal H}_\perp=-\frac{t_\perp}{2}\sum_{i,\ell}\left(b^{\dagger}_{i,\ell}b_{i,\ell+1}^{\dagga}+b^{\dagger}_{i,\ell+1}b_{i,\ell}^{\dagga}\right),
\label{eq:Hhcb2}
\ee
where $\ell$ stands for the layer index,
a true 3D long-ranged-ordered phase is expected at finite temperature, where the KT superfluid-only regime immediately turns into a Bose condensed superfluid with 3D phase coherence. The 3D critical temperature can be estimated from rather simple arguments: If one assumes decoupled superfluid layers, approaching the KT transition from above the correlation length rapidly increases as
\be
\xi_{\rm 2D}(T)\simeq \xi_0\exp\left(b\sqrt{{\frac{T_{\rm{KT}}}{T-T_{\rm{KT}}}}}\right).
\ee
A crossover to the 3D regime is expected when  $t_{\perp}\xi^{2}_{\rm 2D}\ge t_{\parallel}$, i.e. below
\be
T_{\rm 3D}= T_{\rm{KT}}\left(1+\frac{4b^2}{\ln^2(t_{\perp}\xi_0^2/t_\parallel)}\right).
\label{eq:Tc}
\ee
Interestingly, such an estimate turns out to give the correct behavior for the actual $T_c$ at small $t_\perp$, 
as visible in {{Fig.~\ref{fig:PHDGCLEAN}}} where QMC results are displayed for weakly coupled 2D superfluid layers of hard-core bosons at half-filling ($\mu=0$). These results, shown in details in Fig.~\ref{fig:clean}, have been obtained using the SSE algorithm~\cite{Sandvik02} at finite temperature for various 3D arrays of $N=L^3$ sites with $L=8,\cdots,48$. Taking an intralayer hopping strength $t_\parallel=1$, several orders of magnitudes for interlayer Josephson couplings have been explored: $t_\perp=10^{-3},\cdots,1$. 
\begin{table*}
\centering
\begin{tabular}{c||c|c|c|c|c|c|c|c}
${t_\perp}/{t_\parallel}$&0.001&0.005&0.01&0.03&0.05&0.1&0.5&1\\
\hline
${T_c}/{t_\parallel}$&0.395(5)&0.429(5)&0.450(5)&0.490(5)&0.53(1)&0.575(10)&0.85(1)&1.008(3)
\end{tabular}
\caption{{Numerical estimates from Fig.~\ref{fig:PHDGCLEAN} for the 3DXY transition temperature $T_c/t_\parallel$ of clean superfluid layers as a function of the interlayer hopping $t_\perp/t_\parallel$.}}
\label{tab:1}
\end{table*}
%
For weak amplitudes $t_\perp/t_\parallel \le 1/10$, the critical temperature obeys the simple prediction Eq.~\ref{eq:Tc} with {{$T_{\rm{KT}}=0.342 t_\parallel$,}}
$b=1.86$ and $\xi_0=0.34$, as displayed in {{Fig.~\ref{fig:PHDGCLEAN}}}. It is worth noting the strong logarithmic enhancement of the transition temperature above $T_{\rm KT}$ {{(for numerical values, see table~\ref{tab:1}).}}

The superfluid response along all spatial directions can be probed by imposing twisted boundary conditions along a given direction, which translates into the computation of winding number fluctuations in the QMC scheme~\cite{Ceperley87}. The in-plane superfluid stiffness $\rho_{s}^{\parallel}$ is plotted together with the transverse response $\rho_{s}^{\perp}$ in the panels (a) and (b) of Fig.~\ref{fig:clean} for $t_\perp/t_\parallel=1,~0.1,~0.01$. There, one clearly sees that quantum coherence of the superfluid establishes in  both directions, even for very weak $t_\perp/t_\parallel$. When normalized with respect to $t_\parallel$ or $t_\perp$, the superfluid response displays a small anisotropy, only observed for infinitesimal $t_\perp$. 
Importantly, longitudinal and transverse stiffnesses vanish at the same critical temperature $T_c$. This is best visible in Fig.~\ref{fig:clean} (c-d), where the 3DXY transition is detected by using the critical scaling of the stiffness $\rho_s\sim L^{2-D-z}$ where $D=3$ and $z=0$.
Indeed, QMC data of $\rho_s\times L$ display nice crossing features for different system sizes for both components: longitudinal $\rho_s^\parallel$ and transverse $\rho_s^\perp$.\\

\begin{figure*}
\centering
\includegraphics[clip,width=2\columnwidth]{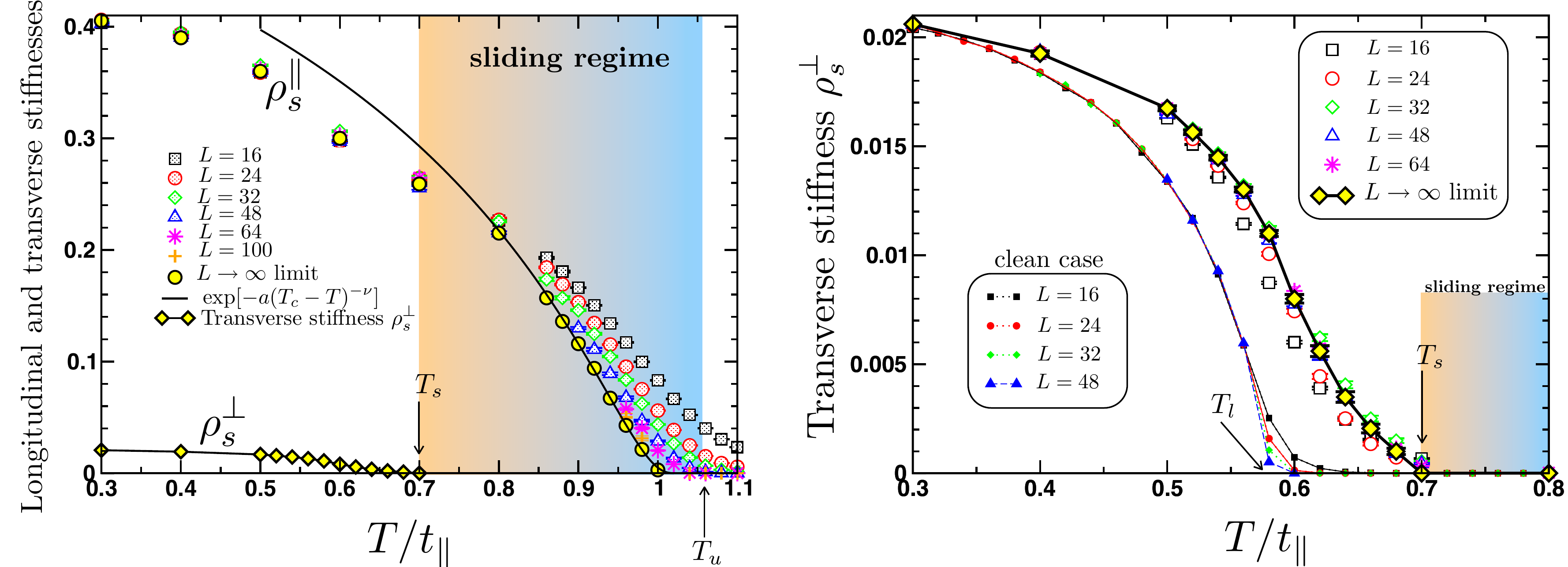}
\caption{(color online) QMC results for the superfluid responses in the longitudinal $\rho_s^\parallel$ (Left) and the transverse $\rho_s^\perp$ (Right) directions plotted against the temperature $T/t_\parallel$ for different 3D systems $L\times L \times L$ of size $L=16,~24,~32,~48,~64,~100$. Disorder averages have been performed from more than $10^4$ samples for $L=16$ to $10^2$ samples for the largest lattices, such that error bars are typically smaller than symbols sizes. When extrapolated to the infinite size limit (yellow symbols), the superfluid stiffness in the plane direction $\rho_s^\parallel$ vanishes for $T_c=1.04(2) t_\parallel$ whereas the response in the transverse direction $\rho_s^\perp$ has already vanished for $T_s\simeq 0.7 t_\parallel$, thus defining a sliding regime (shaded area) for $0.7\le T\le 1.04(2)$. In the right panel, the transverse stiffness for disorder-free A layers with $t_\perp/t_\parallel=1/10$ are also shown, defining $T_l$ (see text).}
\label{fig:sliding}
\end{figure*}

{\it{Randomly stacked layers: sliding regime---}} As predicted in Refs.~\cite{Mohan10,Pekker10}, an unconventional intermediate temperature regime with 2D-only superfluidity can be expected in a disordered layered system, if disorder is introduced in a way that rare regions have  distinct KT temperatures. In order to investigate precisely such a physics, we focus on a simple 3D quantum model of hard-core bosons where $L$ layers of two types (A and B) are randomly stacked and weakly coupled:
\bea
{\cal H}=&-&\sum_{\ell =1}^{L}\Bigl[\frac{t_{\parallel}^{\ell}}{2}\sum_{\langle i j\rangle}\left(b^{\dagger}_{i,\ell}b_{j,\ell}^{\dagga}+b^{\dagger}_{j,\ell}b_{i,\ell}^{\dagga}\right)+\mu\sum_i n_{i,\ell}\nonumber\\
&+&\frac{t_{\perp}}{2}\sum_{i}\left(b^{\dagger}_{i,\ell}b_{i,\ell+1}^{\dagga}+b^{\dagger}_{i,\ell+1}b_{i,\ell}^{\dagga}\right)\Bigr],
\label{eq:5}
\eea
where the in-plane hopping $t_{\parallel}^{\ell}=1$ (A) or 2 (B) with probability $1/2$, and a constant interlayer Josephson tunneling {{$t_\perp=t_\parallel/10$}}. In the following, we fix the chemical potential $\mu=0$ such that the system remains at half-filling. 
Taken independently, each layers exhibit, when decoupled, individual Kosterlitz-Thouless temperatures $T_{\rm KT}^{\rm A}=T_{\rm KT}^{\rm B}/2$. When the transverse coupling $t_\perp$ is turned on, rare thick slabs of $\cal N$ consecutive layers of the same type (A or B) appear with a probability $2^{-\cal N}$. In the infinite system size limit, the existence of such rare regions define an upper temperature 
\be
T_u=T_{\rm{KT}}^{\rm B}\left[1+{4b^2}/{\ln^2(t_{\perp}\xi_0^2/t)}\right]\simeq1.06t_\parallel, 
\ee
and a lower temperature $T_l=T_u/2$. In this temperature range, QMC simulations (Fig.~\ref{fig:sliding}) show clear evidences for an intermediate phase $T_s\le T\le T_u$ where the longitudinal superfluid response $\rho_s^\parallel$ is finite whereas the transverse one $\rho_s^\perp\equiv 0$. Numerical results have been obtained on very large cubic clusters of size $L\times L \times L$, with $L=16,~24,~32,~48,~64,~100$, and averaged over several hundreds of independent disordered configurations of random stacking. Contrary to the clean situation (Fig.~\ref{fig:clean}) where, even for tiny transverse tunneling $t_\perp$, a full 3D coherence was found below a single ordering temperature $T_c$, with a finite superfluid stiffness in {\it all} spatial directions, in the random stacking case the transverse superfluidity becomes non-zero only for $T\le T_s\simeq 0.7t_\parallel$. Below $T_s$ the finite value of $\rho_{s}^{\perp}$ signals a phase locking between layers. In the present situation, a lower bound for $T_s$ (better than $T_l$) is $T_{\rm KT}^{\rm B}${{$\simeq 0.684 t_\parallel$}}, as given by rare events where the effective tunneling between neighboring B-layers $\to 0$. On the other hand,
close to $T_u$, infinite size extrapolations of $\rho_s^\parallel (T)$ are very well described by the analytical prediction of Ref.~\cite{Mohan10} $\sim \exp\left[-a(T_c-T)^{-\nu}\right]$ with $a= 0.6(1)$, $T_c\simeq 1.04(2)$ and $\nu=0.6(1)$.\\

{\it{Scaling of the transverse stiffness---}} In the sliding regime $T_s\le T\le T_c$,
the absence of transverse superfluid response leads to the following finite size scaling~\cite{Mohan10,Vojta}
\be
\rho_s^\perp(L,T)=\frac{1}{L}\frac{\partial^2 E(\varphi_\perp)}{\partial \varphi_\perp^2}\Bigr|_{\varphi_\perp=0} \propto {L^{1-z_\perp(T)}}.
\label{eq:FSSS}
\ee
%
\begin{figure}[!hb]
\centering
\includegraphics[width=\columnwidth,clip]{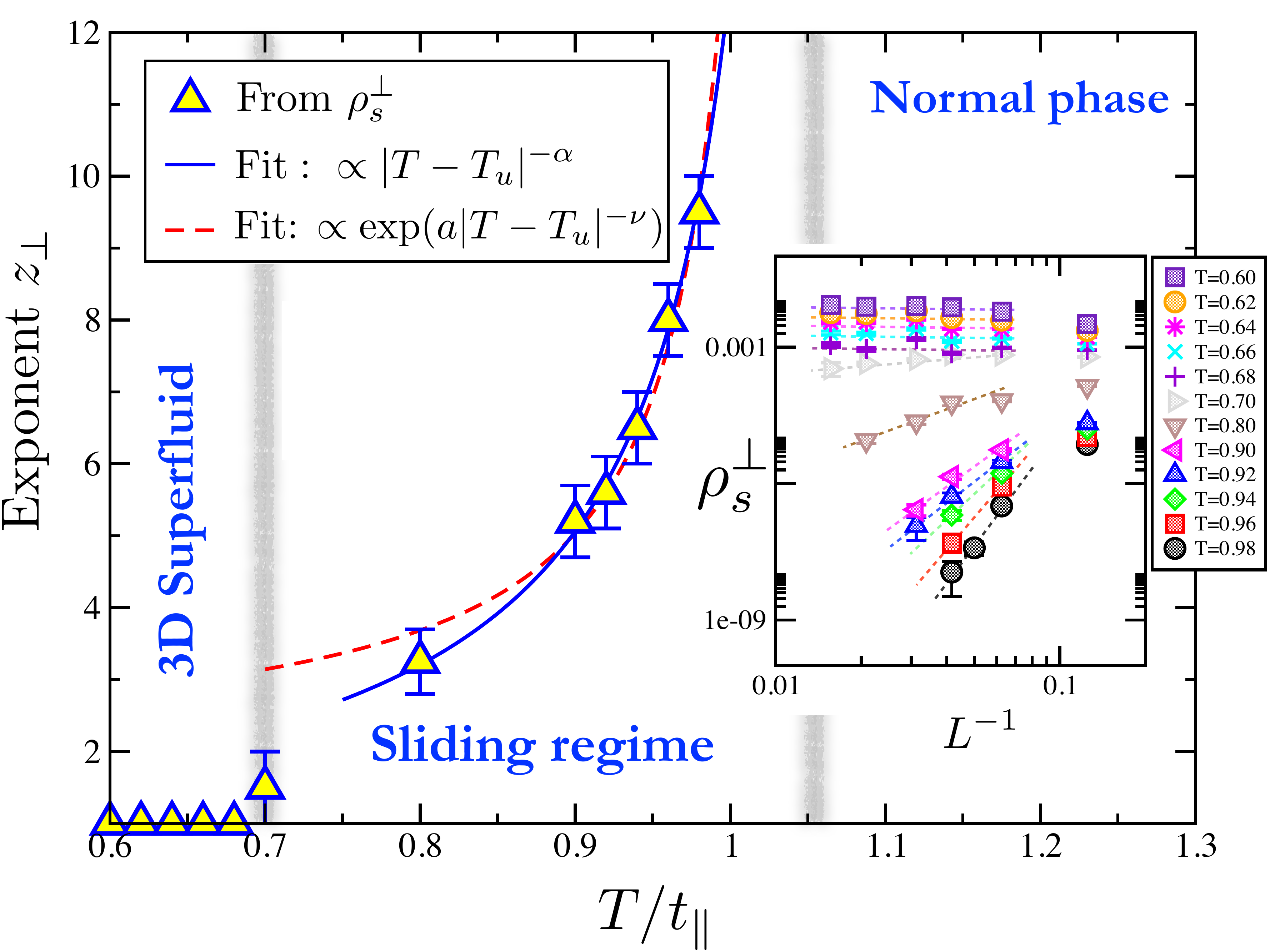}
\caption{(color online) Anomalous exponent $z_\perp(T)$ of the finite size scaling of the transverse stiffness Eq.~\ref{eq:FSSS} estimated from QMC data shown in the inset. The blue line is a fit to  a power-law $\propto |T-T_u|^{-\alpha}$ with $\alpha=0.94$ and the dashed red line is an exponential fit $\propto \exp(a|T-T_u|^{-\nu})$ with $a\simeq 0.33$ and $\nu\simeq 0.67$.}
\label{fig:Z}
\end{figure}

Here $\varphi_\perp$ is a twist angle imposed in the perpendicular direction, and $z_\perp(T)$ is the anomalous exponent 
for  transverse superfluidity. Finite size scaling for $\rho_s^\perp(L,T)$ is displayed in the inset of Fig.~\ref{fig:Z} for various temperatures in the range $0.60\le T/t_\parallel\le 0.98$. In the main panel of Fig.~\ref{fig:Z}, estimates for the anomalous dynamical exponent $z_\perp(T)$ are plotted and display a clear divergence when $T$ approaches $T_u$. \\

A precise fit is not reasonably possible since both power-law and exponential divergences describe correctly the data on such a reduced scale. From Refs.~\cite{Mohan10,Vojta}, we expect a power-law for SU(2) symmetry and exponential in the present U(1) case.\\
\begin{figure}[!h]
\centering
\includegraphics[width=\columnwidth,clip]{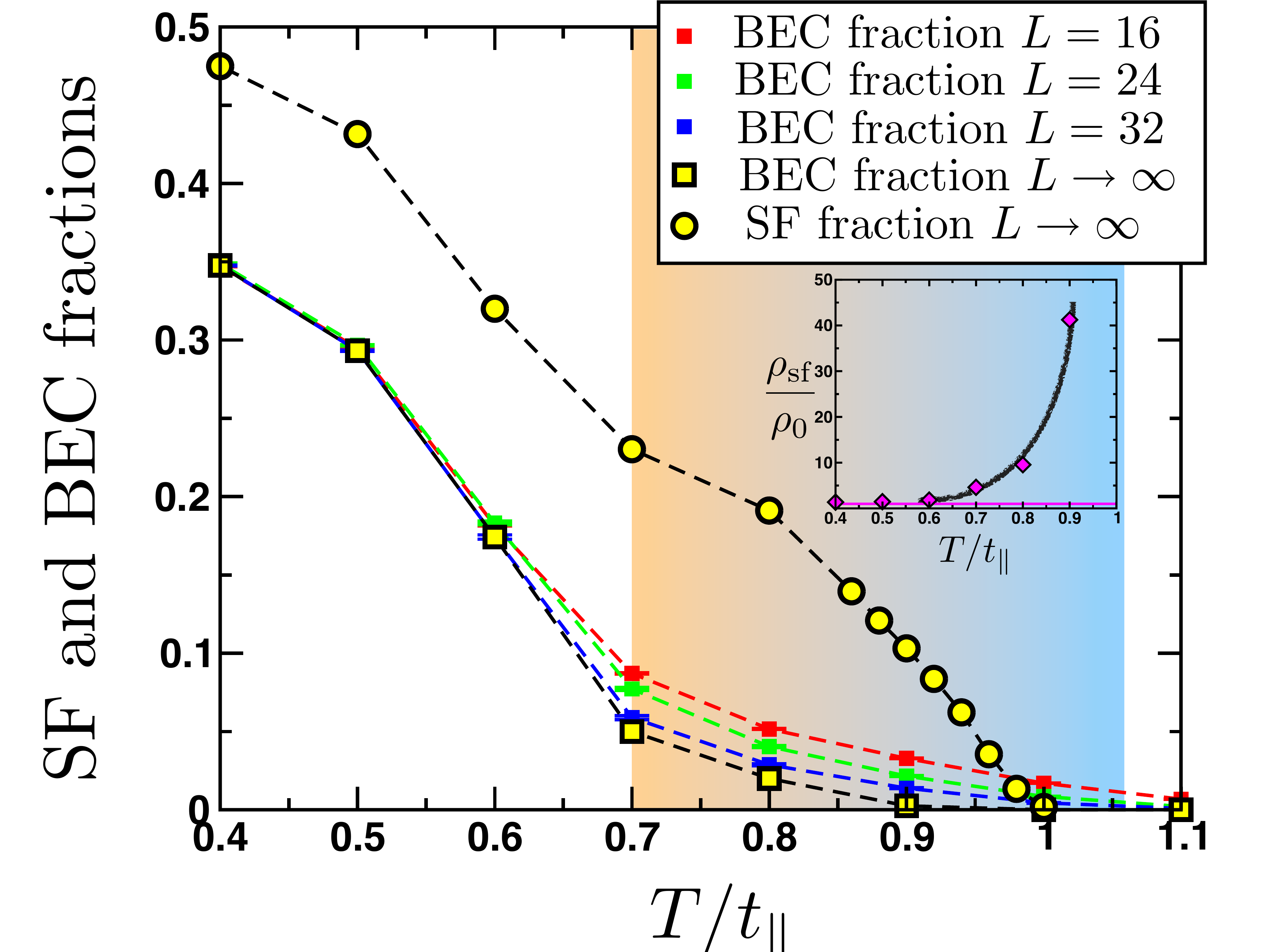}
\caption{(color online) QMC results for SF and BEC fractions of the 3D disordered layered system. Infinite size extrapolations are shown (yellow symbols) for both fractions. Inset: ratio between $L\to\infty$ SF and BEC fractions. The horizontal line is $\rho_{\rm sf}/\rho_0=1$ and the dark curve is a guide to the eyes.}
\label{fig:FRACTIONS}
\end{figure}

{\it{Momentum space properties---}} An interesting issue lies in the momentum space properties. A first observation concerns the relationship between Bose-Einstein condensation and superfluidity. Although the relative phases can fluctuate from one layer to the other in the sliding regime, they remain locked in each plane. Therefore, a macroscopic occupation of the ${\bf k}=0$ mode is expected, as visible in Fig.~\ref{fig:FRACTIONS} where the condensate fraction 
\be{\rho_0}/{\rho}=\frac{1}{N}{\sum_{i,j}\langle b_i^\dagger b_j^\dagga\rangle}/{\sum_{i}\langle b_i^\dagger b_i^\dagga\rangle}\ee
is displayed together with the superfluid fraction (directly obtained from the average over all spatial directions of the superfluid stiffnesses~\cite{Fisher73}).
From a macroscopic point of view, below $T_u$ the disordered layered system displays both superfluidity {\it and} Bose-condensation, without clear evidence for an absence of transverse superfluid response. Nevertheless, inside the sliding regime there is an anomalously small BEC fraction as compared to the SF fraction (see inset of Fig.~\ref{fig:FRACTIONS}), whereas below $T_s$ where full 3D coherence is recovered, both fractions are of the same order of magnitude, as expected for a conventional superfluid.\\

\begin{figure}[!h]
\centering
\includegraphics[width=\columnwidth,clip]{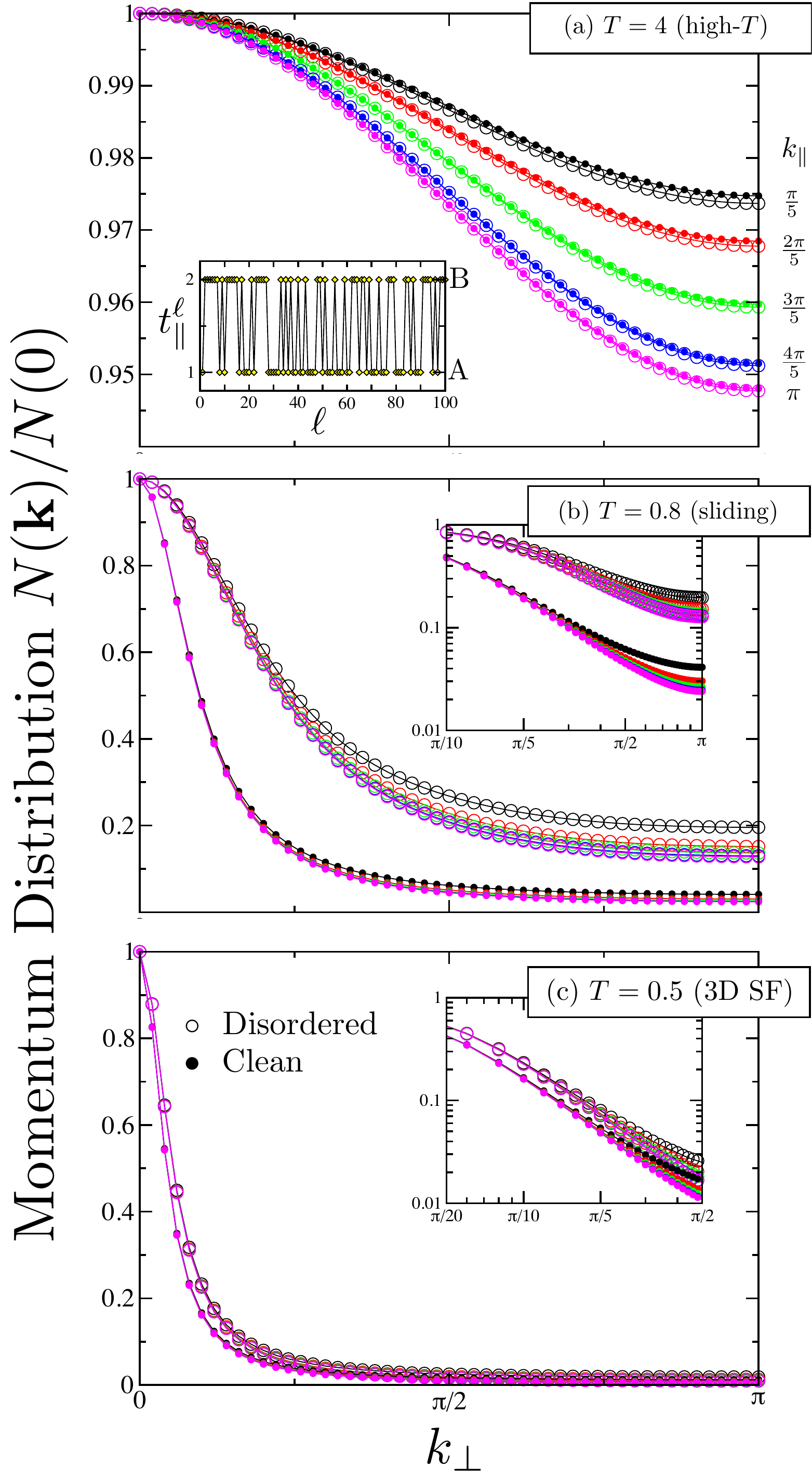}
\caption{(color online) QMC results for the momentum distribution functions of a single random stacking of 100 A and B layers (see inset a) of $10\times 10$ sites, plotted against the transverse momenta $k_\perp$ for various longitudinal wave vectors $k_\parallel$, as shown on the plot. 3 representative temperature regimes are displayed for the disordered system {{($\circ$)}} and are compared with results for a clean sample{{ ($\bullet$)}} with $t_\parallel=2$. }
\label{fig:k}
\end{figure}

The investigation of the momentum space distribution \be N({\bf k})=\sum_{{\bf{r}},{\bf r'}}\langle b_{\bf r}^{\dagger}b_{\bf r'}^{\dagga}\rangle \exp[i{\bf k}\cdot{(\bf r}-{\bf r'})]\ee turns out to be even more instructive. In Fig.~\ref{fig:k}, QMC results for the distribution $N({\bf k})$ of a single disordered sample (a random stacking of $100$ 2D layers of size $10\times 10$) are plotted against $k_{\perp}$, the wave vector in the transverse direction, for 3 representative temperature regimes: (a) normal phase, (b) sliding regime, (c) 3D superfluid. Whereas there is no qualitative difference between clean and disordered systems for (a) and (c), the sliding regime (which is similar to (c) in the clean case) features a much broader momentum distribution in the disordered case. We believe that such a qualitative feature would be observable using time of flight imaging in cold atom experiments.\\
\\

\noindent{\it{Conclusions---}}
{{Before concluding, we may briefly discuss the generality of our results. A schematic tentative phase diagram is given in Fig.~\ref{fig:final} in the $t_{\perp}/t_{\parallel}^{\rm A} - T/t_\parallel^{\rm A}$ plane for model (\ref{eq:5}). The key ingredient to observe the sliding regime, valid for other kinds of random stacking or disorder, lies in the existence of a finite temperature window $T_{s}< T < T_{u}$ where exponentially rare regions of infinite extend having the smallest ordering temperature remain disordered whereas other layers are superfluid.\\

{{
\begin{figure}
\centering
\includegraphics[width=\columnwidth,clip]{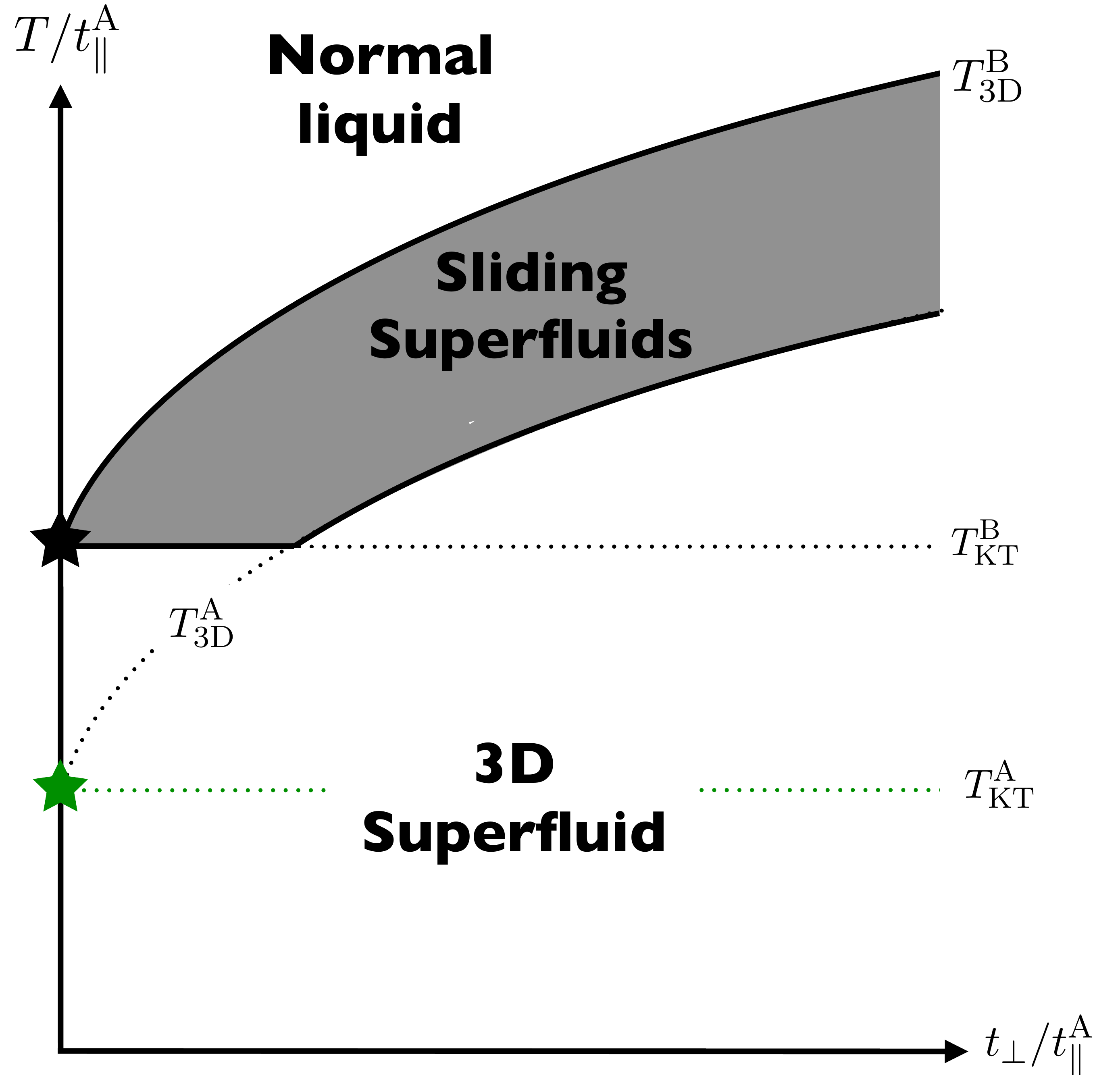}
\caption{{{(color online) Schematic phase diagram for randomly stacked superfluid layers of type A or B [Eq.~(\ref{eq:5})].}}}
\label{fig:final}
\end{figure}
}}
To conclude,}}
we have numerically studied the effect of a weak 3D Josephson tunneling between superfluid or superconducting layers. Contrary to the case of clean identical planes, when distinct layers are randomly stacked an intermediate sliding regime emerges over a finite temperature window, as predicted recently~\cite{Mohan10,Pekker10}. The absence of 3D phase locking is accompanied with a diverging exponent $z_\perp (T\to T_c)$, a finite but very small BEC fraction, and a broadening of the momentum distribution $N({\bf k})$.\\
\\
\acknowledgments
I thank F. Mila and T. Vojta for usefull comments and discussions.

\end{document}